\documentclass[twocolumn,floats,showpacs,preprintnumbers,amsmath,amssymb,epsf,prb]{revtex4}
\usepackage{graphicx}
\usepackage{dcolumn}
\usepackage{bm}
\begin{document}
\title{Low-temperature phases in PbZr$_{0.52}$Ti$_{0.48}$O$_{3}$ : A neutron powder diffraction
study}
\author{D. E. Cox}
\affiliation{Physics Department, Brookhaven National Laboratory, Upton, NY
11973, USA}
\author{B. Noheda}
\altaffiliation[Present address: ]{Materials Science Center, University of Groningen, Nijenborgh 4, 9747 AG Groningen, The Netherlands} \affiliation{Physics Department, Brookhaven
National Laboratory, Upton, NY 11973, USA}
\author{G. Shirane}
\affiliation{Physics Department, Brookhaven National Laboratory, Upton, NY
11973, USA}

\date{July 21, 2004}
\begin{abstract}
A neutron powder diffraction study has been carried out on PbZr$_{0.52}$Ti$_{0.48}$O$_{3}$
in order to resolve an ongoing controversy about the nature of the
low-temperature structure of this strongly-piezoelectric and
technologically-important material. The results of a detailed and systematic
Rietveld analysis at 20 K are consistent with the coexistence of two
monoclinic phases having space groups $Cm$ and $Ic$ respectively, in the
approximate ratio 4:1, and thus support the findings of a recent electron
diffraction study by Noheda {\it et al.} [Phys. Rev. B {\bf  66}, 060103 (2002)]. The
results are compared to those of two recent conflicting neutron powder
diffraction studies of materials of the same nominal composition by Hatch {\it et
al.} [Phys. Rev. B {\bf 65}, 212101 (2002)] and Frantti {\it et al.} [Phys. Rev. B {\bf 66},
064108 (2002)].
\end{abstract}
 \maketitle

\section{Introduction}
The strongly piezoelectric system PbZr$_{1-x}$Ti$_{x}$O$_{3}$ (PZT) has long
been known to have a perovskite-type structure with regions of rhombohedral
and tetragonal symmetry below the ferroelectric Curie point separated by an
almost vertical line at $x \approx 0.5$ in the temperature-composition phase
diagram, which is known as the morphotropic phase boundary (MPB).\cite{Jaffe}%
 Following the recent discovery of a narrow region with monoclinic $Cm$
symmetry in the vicinity of the MPB,\cite{Noheda1} numerous experimental
and theoretical studies of PZT and related systems have been undertaken in
order to clarify the relationships between the structural features and the
piezoelectric properties. As previously discussed,\cite{Noheda2} the
ferroelectric polarization in the new phase is no longer constrained by
symmetry to lie along a symmetry axis, but instead is free to rotate within
the symmetry plane. Furthermore, because of the near-degeneracy of the free
energies of the various phases, rotation of the polarization axis away from
the polar axes of the rhombohedral and tetragonal phases can be accomplished
with an applied electric field, resulting in an induced monoclinic phase and
a large electromechanical response.\cite{Bellaiche,Cohen}

The phase diagram of the PZT system around the MPB as reported in a recent
paper by Noheda {\it et al.}\cite{Noheda3} is shown in Fig. 1. Above the Curie
temperature, the structure is cubic over the entire range of composition,
with space group $Pm3m$ and lattice parameter $a_{0}\approx ${} 4 \AA {}. The
rhombohedral region is characterized by high- and low-temperature phases ($\rm R%
_{HT}$ and $\rm R_{LT}$) in which there are polar shifts of the
atoms along the pseudocubic [111] axis.\cite{Michel,Corker} $\rm
R_{HT}$ has space group symmetry $R3m$, with lattice parameters
$a_{R} \approx  a_{0}$, and $\alpha $ slightly less than 90$^{\circ}$
(hexagonal values $a_{H} \approx a_{0}\sqrt{2}, {} c_{H} \approx
a_{0}\sqrt{3}$). In $\rm R_{LT}$ there are additional displacements
of the oxygen atoms superimposed on the ferroelectric shifts due
to antiphase tilting of the oxygen octahedra about the [111] axis,
corresponding to an R-point instability. As a consequence, the
unit cell is doubled and the mirror plane is destroyed, resulting
in the appearance of superlattice peaks in the diffraction
pattern. The new space group symmetry is $R3c$, with hexagonal
lattice parameters $a_{H} \approx a_{0}\sqrt{2}$ and $c_{H}\approx
2a_{0}\sqrt{3}$. It should be emphasized that these must be
regarded as \lq\lq average'' long-range structures, since the
presence of short-range order due to local displacements has been
clearly demonstrated by the appearance of other types of
superlattice peaks in electron diffraction
studies\cite{Viehland1,Viehland2,Ricote} not observed in x-ray or
neutron diffraction patterns.\cite{Michel,Corker} Significant
deviations of the local atomic structure from the crystallographic
long-range structure have also been found from pair-distribution
function (PDF) analysis of time-of-flight neutron
data.\cite{Teslic1,Teslic2,Dmowski}

In the tetragonal region of the phase diagram, the space group is
$P4mm$ and the polar shifts lie along the [001] axis ($a_{T}
\approx c_{T} \approx a_{0},{} c_{T}/a_{T} > 1)$. Nevertheless,
the time-of-flight neutron data show that this too should be
viewed as an \lq\lq average'' long-range structure. In addition,
Raman scattering studies have revealed the presence of local
displacements of lower symmetry, which are also reflected in a
broadening of some of the x-ray diffraction
peaks.\cite{Frantti1,Frantti2} The nature of the local structure
has been revealed in more detail  from the PDF analysis described
in Ref. 14, which shows that there are only gradual changes
through the MPB, and suggests that the local environment of each
element remains relatively invariant of composition. It is
furthermore proposed that the population of local Pb displacements
changes between the pseudocubic $<$100$>$ and $<$110$>$ directions
as a function of the Ti/Zr ratio. This model is supported by
recent theoretical calculations in which the Pb distortions are
identified as the determining factor for the average structure of
the system.\cite{Grinberg}

 In the original x-ray
study by Noheda {\it et al.}\cite{Noheda1} the unit cell of the low-temperature
monoclinic phase (now usually designated $\rm M_{A}$\cite{Cohen}) was found to be doubled
with respect to the primitive cell, with the monoclinic $a$ and $b$ axes
directed along the [110] and [1$\overline{1}$0] axes of the latter ($a%
_{M} \approx b_{M} \approx a_{0}\surd 2, c_{M} \approx a_{0}$,
space group $Cm$). Based upon the atomic positions determined from
Rietveld analysis of the synchrotron x-ray data from PbZr
$_{0.52}$Ti$_{0.48}$O$_{3}$, it was concluded\cite{Noheda2} that
at 20 K the polar axis was tilted about 24$^{\circ}$ from the [001] axis
towards the pseudocubic [111] axis. The structure can be regarded
as a condensation of either the local displacements present in the
tetragonal $P4mm$ phase along one of the $\left\langle
110\right\rangle $ directions, or alternatively those present in
the rhombohedral $R3m$ phase along one of the $\left\langle
100\right\rangle $ directions, as inferred by Corker {\it et
al.}\cite{Corker}

 However, it is clear that there is a missing
ingredient in this simple picture, for in a neutron powder
diffraction study of the same sample, Noheda {\it et al.} reported
the presence of one very weak superlattice peak at 20 K
corresponding to a doubling of the $c$ axis of this monoclinic
cell, but did not identify the nature of this additional
distortion.\cite{Noheda3} A similar cell-doubled phase was also
observed for $x = 0.48$ by Ragini and coworkers in electron
diffraction patterns below 200 K, but not in their low-temperature
x-ray patterns.\cite{Ragini} Based on a subsequent Rietveld
analysis of neutron powder data collected at 10 K, the structure
of this new phase was reported by Ranjan {\it et al.} to be
monoclinic, with space group $Pc$.\cite{Rajan} The appearance of
the weak superlattice reflections was attributed to antiphase
tilting of the oxygen octahedra about the [001] direction,
corresponding to an R-point instability in the cubic Brillouin
zone. It was later reported that the correct space group for this
proposed model was in fact $Cc$,\cite{Hatch} and a modified set of
refined structural parameters was presented.

The $x = 0.48$ composition has
also been the subject of a recent low-temperature neutron powder study by
Frantti and colleagues.\cite{Frantti3} They, too, note the presence of
similar superlattice reflections, but reach very different conclusions;
namely that these reflections are attributable to a minority rhombohedral
phase with $R3c$  symmetry in coexistence with the monoclinic $Cm$  phase, a model
that was not considered by Ranjan {\it et al.}\cite{Rajan} or Hatch {\it et al.}\cite
{Hatch} In a footnote to their paper, Frantti {\it et al.} comment that the
monoclinic $ Pc$  and $Cc$  models proposed by the latter authors predict peaks
that are not observed experimentally, and that the observed superlattice
peaks can be accounted for by the $R3c$  phase. However, this conclusion was
not supported by the results obtained by Noheda {\it et al.}\cite{Noheda4} in an
electron diffraction study of the same $x = 0.48$ sample used in the earlier
x-ray study,\cite{Noheda2} which showed no evidence for a rhombohedral
phase, but instead the monoclinic $Cm$ phase in coexistence with nanoregions
of a minority $Cc$ phase ranging in size from 30-100 \AA {}. These conclusions
have been questioned by Frantti {\it et al.}, who comment that their neutron data
provide no evidence of a $Cc$ phase, and argue that since electron diffraction
probes only small volumes of the sample, it is generally not suitable for
the determination of average symmetry, and furthermore that the ion-milling
technique used for sample thinning is a very violent one which can easily
generate significant defects.

In the light of these different interpretations, we have undertaken a
detailed Rietveld analysis of the neutron data cited by Noheda {\it et al.}\cite
{Noheda3} in an attempt to discriminate between the three models described
above. Plausible results were obtained in all three cases, illustrating how
difficult it is to identify the correct structural model in complex systems
of this type simply on the basis of the standard goodness-of-fit criteria.
Nevertheless, we conclude that, taken in conjunction with the electron
diffraction data, the results point strongly towards the coexistence model
of  $Cm$ and minority $Cc$ phases.
\section{Experimental}
The sample consisted of about 4 g of sintered pellets roughly 1 cm in
diameter and 1 mm thick from the same batch of material used in the previous
x-ray study. \cite{Noheda2} Long-range fluctuations in the composition of
the x-ray sample, $\Delta x$, were estimated to be less than $\pm $ 0.003
based upon an analysis of the peak widths. The pellets were loaded into a
thin-walled vanadium can and mounted in a closed-cycle helium cryostat. Data
were collected at the NIST reactor on the powder diffractometer BT1 with a
Cu monochromator set for a wavelength of 1.54 \AA {}, collimation of 15$^{\prime}$ and
40$^{\prime}$ before and after the monochromator, and 10$^{\prime}$ in front of each of the 32 $%
^{3}$He detectors. With this configuration, the best angular
resolution attained is about 0.2$^{\circ}$ at 2$\theta \approx$ 80$^{\circ}$
($\Delta d/d
\approx 2 \times 10^{-3}$).

Extended data sets were collected at 2$%
\theta $ step intervals of 0.05$^{\circ}$ in the monoclinic region at 20
K, in the vicinity of the monoclinic-tetragonal transition at 325
K, and in the tetragonal region at 550 K. Analysis of the data was
carried out with the FULLPROF program,\cite{FULLPROF} using the
pseudo-Voigt peak-shape function with appropriate corrections for
instrumental asymmetric broadening, \cite {Finger} and linear
interpolation between background points. Particular attention was
paid to the problem of anisotropic peak-broadening, which reflects
the fact that closely-adjacent peaks may have markedly different
widths arising from local strains or compositional fluctuations,
for example, as previously noted for PZT and related piezoelectric
systems.\cite {Noheda1,Noheda3,Frantti2,Kiat}  In standard
Rietveld analysis the peak widths are assumed to vary smoothly as
a function of scattering angle, and it is important to note that
anisotropic peak-broadening due to microstructural effects can be
mistakenly interpreted as a symmetry-lowering distortion of the
unit cell of the average long-range structure. With the rapidly
increasing use of high-resolution x-ray and neutron techniques, it
is becoming clear that anisotropic peak-broadening is a common
feature of powder diffraction patterns, and should be allowed for
as appropriate. One convenient way to do this is provided by the
phenomenological model recently proposed by
Stephens,\cite{Stephens} in which the broadening is represented by
a series of coefficients $\sum_{HKL}S_{HKL}$$h^{H}k^{K}l^{L}$
$(H+K+L=4)$, which take into acccount the Laue symmetry of the
space group and are incorporated as refinable parameters in the
Rietveld program. For tetragonal $4mm$ and monoclinic $2/m$
symmetry, there are respectively 4 and 9 such coefficients.

The data analysis is now described in detail for the tetragonal phase at 550
K, the monoclinic phase at 20 K, and the intermediate region at 325 K.
\subsection{550 K}
All the peaks could be unambiguously indexed in terms of a tetragonal cell
with $a = 4.060, c = 4.100$ \AA {}, except for two very weak peaks
attributable to the vanadium sample holder, which were excluded from the
analysis. Rietveld refinement was carried out with individual isotropic
temperature factors assigned and the atoms placed in the following positions
of space group $P4mm$: Zr/Ti and O(1) in 1(a) sites at $0, 0, z$; O(2) in 2(c)
sites at $0.5, 0, z$; and Pb statistically distributed among the 4(d) sites at
$x, x, 0$. The Pb positions correspond to random displacements in the
$\left\langle 110\right\rangle$ directions away from the origin, as noted
 in the previous synchrotron x-ray study.\cite{Noheda2}  The
refinement converged rapidly and smoothly to a goodness-of-fit $\chi ^{2}$
value of 1.30. However, as  in the x-ray study, an examination
of the observed and calculated peak profiles revealed a number of systematic
discrepancies indicative of anisotropic peak broadening, and additional
refinements were carried out in which various combinations of the four
possible coefficients were allowed to vary. A definite improvement was
obtained when $S_{004}$ was varied alone ($\chi ^{2}$ = 1.20), but the
results obtained with additional coefficients were judged to be of dubious
significance, and these coefficients were accordingly set to zero. This
result most likely reflects the sensitivity of the $c$ lattice parameter to
the presence of long-range compositional fluctuations. The final refinement
results based on this model are listed in Table I, and the profile and
difference plots are shown in Fig. 2.
\subsection{20 K}
A series of refinements was carried out for each of the three models
described in the Introduction, namely: single-phase $Cc$, two-phase $Cm/Cc$,
and two-phase $Cm/R3c$. However, instead of  $Cc$, the non-conventional space
group setting $Ic$ was chosen, which has the distinct advantage of having a
near-orthogonal unit cell closely related to the $Cm$ cell, in which the
mirror plane is replaced by a $c$-glide plane and the $c$ axis is doubled. The
unit-cell axes are related via the transformation {\bf a}$_I$ = -{\bf c}$_C$,
{\bf b}$_I$ = {\bf b}$_C$, {\bf c}$_I$ = {\bf a}$_C$ + {\bf c}$_C$, where the
subscripts refer to the unit cells of the $Ic$ and $Cc$ space groups respectively. In
this setting, it is much easier to visualize the small displacements from
the ideal $Cm$ atomic positions. In the $Ic$ cell, the Pb atom was chosen to lie
at the origin, with Zr/Ti and three inequivalent O atoms in fourfold general
positions at $x, y, z$ and $x, -y, 1/2+z$, and at the related body-center sites.
The Zr/Ti and O(1) atoms are in positions similar to those in the $Cm$
structure at $x, 0, z$, the main difference being that they are no longer
required to lie on a mirror plane at $y = 0$. The O(2) and O(3) atoms are in
two sets of  positions derived from the $x, y, z$ sites and the symmetry-equivalent
mirror plane sites at $x, -y, z$ in the $Cm$ structure.
\subsubsection {Single-phase $Ic$ model}
 In the first series of refinements the
atoms were intially assigned the positions found in our previous
x-ray study\cite {Noheda2} with the exception of the O(2) and O(3)
atoms, which were displaced from the ideal $Cm$ positions by small
shifts corresponding to antiphase tilting of the oxygen octahedra
about the [001] axis, as assumed by Ranjan {\it et
al.}\cite{Rajan} The corresponding positions chosen for O(2) and
O(3) in the $Ic$ structure were $x - \delta, y - \delta, z/2$ and
$1/2 + x + \delta, 1/2 + y + \delta, z/2$, where $\delta$ is the
shift in the $x$ and $y$ directions due to tilting. It is
important to note that with such a constrained-tilt,
rigid-octahedron model, the $x$ and $y$ values assumed for the
O(2) and O(3) positions (in this case the $Cm$ values found in the
previous x-ray study) do not change in the course of the
refinement. The resulting fit was reasonably good ($\chi ^{2}$ =
1.97), but inspection of the individual peak profiles once again
revealed some significant discrepancies due to anisotropic peak
broadening. As before, a distinct improvement was obtained when
the anisotropy coefficient $S_{004}$ was refined ($\chi ^{2}$ =
1.78), but further refinements with various combinations of the
other eight anisotropy coefficients gave only minimally improved
fits, and the one-parameter anisotropy model was accordingly
adopted for subsequent refinements.
At this point, the constraints on the Zr/Ti and O(1) $y$ parameters were
relaxed, but the shifts from the ideal positions at $y$ = 0 and the
improvement in the overall fit were judged to be insignificant. A similar
result was obtained when the Zr/Ti compositional parameter $x$ was varied.

Further refinements were performed, first with constrained models
corresponding to rigid-octahedron tilting about the [111] and [110] axes
respectively, and finally, with all the constraints on the O(2) and O(3)
positions removed. In the latter case, the refinement proceeded smoothly and
converged rapidly to a set of positions which were much closer to those of
the [001]-tilt model than the other tilt models independent of which tilt
model was used to provide the initial values of the positions. However, in
none of these cases did the overall fit appear to be significantly improved,
and we therefore conclude that the [001]-tilt model is a reasonable choice,
although it is clearly not possible to rule out the other models on the
basis of the present data. The final refinement was therefore carried out
for the constrained [001]-tilt model, but with the $x$ and $y$ values for O(2)
and O(3) derived from the results for the unconstrained model. This
refinement yielded a $\chi ^{2}$ value of 1.52, with refined parameters as
listed in Table II (column 1). Also shown are the values reported by Hatch
{\it et al.}\cite{Hatch} transformed from $Cc$ to $Ic$ symmetry (column 3). From a
comparison of the two sets of atomic positions, it appears that the
constraints applied by Hatch {\it et al.} do not in fact correspond to an
[001]-tilt model, but instead to a simpler model in which only the $y$
parameters of the O(2) and O(3) atoms are displaced from their ideal $Cm$
positions. An additional refinement based on such a $y$-shift model yielded
results which are seen to be in excellent agreement with those of Hatch {\it et
al.} (column 2 of Table II), although the fit is somewhat inferior to that
given by the [001]-tilt model ($\chi ^{2}$ = 1.62). It is also worth noting
that the values of axial ratio, $c_{0}/a_{0}$, and the pseudocubic cell volume, $V_{0}$,
obtained by the latter authors suggest a slightly higher Zr content (%
$\approx  0.5$\%) relative to the present sample.\cite{Noheda3,Frantti3}

From the atomic positions listed in the first column of Table II
the octahedral tilt angle is calculated as about 3$^{\circ}$. The
polar displacements of the Zr/Ti and Pb atoms with respect to the
respective polyhedra centers are -0.08 and -0.22 \AA {} along
monoclinic [100], and 0.18 and 0.44 \AA {} along [001],
corresponding to a rotation of the polar axis towards pseudocubic
[111] of roughly 25$^{\circ}$. However, these values are
representative only of the average long-range structure, since
they do not allow for the local distortions revealed in the PDF
analysis cited earlier.\cite{Dmowski}
\subsubsection{Two-phase $Cm/R3c$ model}
The next set of refinements was performed for the two-phase
$Cm/R3c$ model favored by Frantti {\it et al.}\cite{Frantti3}
Significantly better peak profiles were obtained with an
anisotropic-broadening model for the $Cm$ phase in which $S_{004}$
was allowed to vary, together with an isotropic particle-size
broadening coefficient for the $R3c$ phase. The refinement
converged rapidly to a $\chi ^{2}$ value of 1.27 with the final
parameters as listed in Table III. The latter are in close
agreement with those reported by Frantti {\it et al.}, including
the respective weight fractions of the two phases. Compared to the
single-phase $Ic$ model, the overall fit is considerably better
($\chi ^{2}$ = 1.27 vs. 1.52), but because several additional
variable parameters are involved, it is difficult to judge the
true significance of this result.
Inspection of the results listed in Table III reveals that in both samples, $V_{0}$
for the rhombohedral phase is larger by about 0.3 \AA $^{3}$, which
would imply a significantly higher Zr content of some 3-4\%.\cite
{Noheda3,Frantti3}  For the present sample, at least, such a conclusion
would be inconsistent with the previously estimated long-range compositional
fluctuations.\cite{Noheda2}
\subsubsection{Two-phase $Cm/Ic$ model}
The final set of refinements was carried out for the two-phase
$Cm/Ic$ model deduced by Noheda {\it et al.} from the results of
an electron diffraction study.\cite{Noheda4}   Since we did not
anticipate that a meaningful result would be obtained for an
unconstrained refinement of two such closely-related structures, a
highly constrained model was used; namely, the atomic positions in
the $Cm$ and $Ic$ phases were constrained to be equivalent except
for one additional parameter $\delta $ for the latter representing
the displacement along the $x$ and $y$ axes for the idealized
[001]-tilt model described above. The peak-shape model, including
an
 $S_{004}$ anisotropy coefficient, was also constrained to be equivalent for
both phases, except for an isotropic particle-size broadening coefficient
which was included for the $Ic$ phase.
The refined values for the two phases
are listed side-by-side in Table IV for easy comparison, and the profile fit
and difference plot are shown in Fig. 3. The relative proportions of the $Cm$
and $Ic$ phases are approximately 4:1 and thus consistent with the electron
diffraction results, but the estimated particle size derived from the
broadening coefficient is much larger, about 1000 \AA {} compared to
100 \AA {}. As pointed out by Frantti {\it et al.},\cite{Frantti3} this
discrepancy could arise because of the ion-milling techniques used to thin
the electron diffraction sample, which can generate significant numbers of
defects.

Detailed comparison of the results in Tables II, III and IV
reveals that a better fit is obtained with the two-phase $Cm/Ic$
model ($\chi ^{2}$ = 1.16, $R_{wp} = 0.064)$ than with the
$Cm/R3c$ model ($\chi ^{2}$ = 1.27, $R_{wp} = 0.067$) or the
single-phase $Ic$ model ($\chi ^{2}$ = 1.52, $R_{wp} = 0.073$),
but it would nevertheless be premature to conclude that the former
must therefore be correct, since there are no generally-accepted
statistical tests to judge the true significance of the results.
However, although the extended profile and difference plots for
the two latter models are hardly distinguishable by eye from those
shown in Fig. 3, there are significant differences in some of the
individual peak profiles which provide additional insight, as
shown in Figs. 4 and 5 respectively. Fig. 4 shows the region
around the strongest superlattice peak at 2$\theta \approx
36.8^{\circ}$, from which it is evident that a much better fit is
obtained with the $Ic$ and $Cm/Ic$ models than with the $Cm/R3c$
model. On the other hand, the fit shown in Fig. 5 in the
pseudocubic (200) region reveals serious deficiencies for the $Ic$
model compared to the $Cm/R3c$ and particularly the $Cm/Ic$ model,
which accounts much better for the asymmetry of the profiles in
the central region.

We note also that further analysis of the synchrotron x-ray data reported in
Ref. 3 shows that the $Cm/Ic$ coexistence model is superior to
the two-phase $Cm/Pm3m$ model previously used, with $\chi ^{2}$ values of 7.3
and 12.8 respectively. The results are in reasonable agreement with those of
the neutron study; in particular, the ratio of the two phases is found to
be about 4:1, very similar to the value listed in Table IV. The failure to
detect any superlattice peaks analogous to the one in the neutron pattern
can be explained by the relatively much weaker x-ray scattering power of
oxygen compared to Pb and Zr/Ti, resulting in calculated intensities that
are insignificant compared to the background signal.

From the atomic positions listed in Table IV, the octahedral tilt
angle about the [001] axis is calculated as about 7$^{\circ}$. The
polar shifts of the Zr/Ti and Pb atoms with respect to the
polyhedra centers are essentially the same as those obtained for
the single-phase $Ic$ refinement. It is also seen that the values
of the cell volume $V_0$
 and the axial ratio $c_0/a_0$ for the $Ic$ phase are respectively slightly larger and smaller
 than those for $Cm$, and thus suggestive of a slightly higher Zr content ($\approx  0.5\%$)
for the former.\cite{Noheda3,Frantti3}
\subsection{325 K}
Refinement was first carried out based on a model similar to that used for
the 550 K data; namely, a single-phase tetragonal structure with $P4mm$
symmetry, Pb atoms statistically distributed among the 4(d) sites at $x, x,
0$, and a single anisotropy-broadening coefficient, $S_{004}$. However, the
overall fit was only mediocre ( $\chi ^{2}$ = 2.94, $R_{wp} = 0.072$), and a
detailed inspection of the individual peak profiles revealed asymmetries
consistent with the presence of a monoclinic component. Such a coexistence
model of monoclinic and tetragonal phases for $x = 0.48$ at room temperature
was proposed in an earlier neutron study by Frantti {\it et al.},\cite{Frantti4}
and in a more recent x-ray study by Ragini {\it et al.}\cite{Ragini2}  Further
refinements based on this model gave a markedly improved fit ( $\chi ^{2}$ = 1.76,
$R_{wp}$ = 0.056), but some residual diffuse scattering was clearly
present between some of the peaks. This scattering is probably associated
with locally disordered regions in the vicinity of domain walls and can be
modeled in a simple, albeit rather artificial, way by the addition of a
cubic phase with $Pm3m$ symmetry, as assumed in our previous x-ray study.\cite
{Noheda2}  Such a three-phase model yielded a reasonably satisfactory fit
 ($\chi ^{2}$ = 1.47, $R_{wp} = 0.051$), with weight fractions of tetragonal,
monoclinic and cubic phases in the ratio 0.61:0.33:0.06.

The refined parameters are listed in Table V, and the profile fit and difference plot
are shown in Fig. 6. Also listed are the parameters reported by Ragini {\it et
al.}\cite{Ragini} and Frantti {\it et al.}\cite{Frantti3}  In the latter case, it
is rather surprising in the light of the results reported in Ref. 6
 that the  lattice strain $c_{0}/a_{0}$ was found to be
significantly larger for the monoclinic phase than for the tetragonal one, since one would not expect
rotation of the polarization direction away from [001] in the monoclinic
phase to increase this strain. Other than this, the three sets of parameters
are in reasonable agreement except that the fraction of  $Cm$ phase in the room
temperature studies is considerably larger than at 325 K, as would be
expected.

Further analysis of the earlier x-ray data\cite{Noheda2} revealed that this
three-phase model gives a noticeably better profile fit than that obtained
with the two-phase $P4mm/Pm3m$ model previously used, with $\chi ^{2}$ values
of 7.5 and 9.9 respectively. The weight fractions of the three phases were
in the ratio 0.55:0.40:0.05, comparable to the neutron values listed in
Table V. We emphasize, however, that these results should be regarded as
representative only of an average long-range structure, since the
true nature of the material in the transition region is surely far more
complex than implied by a simple three-phase model. It is more likely
in this temperature interval that small fluctuations in
composition lead to coexistence of the tetragonal phase with
locally-ordered monoclinic regions of widely varying sizes and possibly some
disordered regions.
\section{Discussion}
In summary, the results obtained in the present neutron investigation are
consistent with the coexistence of majority $Cm$ and minority $Ic$ phases in PbZr%
$_{0.52}$Ti$_{0.48}$O$_{3}$, in agreement with the results of a recent
electron diffraction study of the same sample. The $Cm$ ($\rm M_{A}$) phase, which
plays a key role in the piezoelectric and ferroelectric behavior of  PZT and
related systems is the majority phase at low temperature. The structure of
the minority $Ic$ phase is readily visualized as the superposition of an
antiphase octahedral-tilt system on the parent $Cm$ structure. Furthermore,
the close agreement between the refinement results for the alternative $Ic$
and $Cm/R3c$ models and those in Refs. 20 and 21
respectively suggests that the two-phase $Cm/Ic$ model is worth consideration
in those cases as well. It is possible that the coexistence of  $Cm$ and $Ic$
phases in the present $x = 0.48$ sample reflects the existence of a narrow
thermodynamically-stable region with $Ic$ symmetry at low temperature
 somewhere between 0.45$ < x <$ 0.48.
In this case, the coexistence of  $Ic$ and $Cm$ phases could plausibly be
attributed to the presence of long-range compositional fluctuations, as suggested
by the values of the lattice parameters for the two-phase refinement in Table V.
In this context, it is interesting to note that recent neutron data obtained by Frantti {\it et al.}
 for a sample with $x = 0.46$ show clear evidence of a superlattice peak at 4 K.\cite{Frantti5}
However, the authors interpret this as evidence for the coexistence of $Cm$ and $R3c$
phases, and did not consider the possibility of $Cm$ and $Ic$ phase coexistence.
Alternatively, the $Ic$ phase could be a metastable one resulting from the
presence of local strains at domain-wall boundaries, for example. Indeed, it
is noteworthy that first-principles calculations by Fornari and Singh have
shown that local stress fields may lead to the coexistence of both
ferroelectric and rotational instabilities near the MPB.\cite{Fornari}  In
any case, it is clear that a very careful high-resolution x-ray, neutron and
electron diffraction study of extremely well-characterized samples would be
required in order to throw further light on these issues.

On a final note of
caution, the present study also demonstrates that the interpretation of the
results of Rietveld analysis may be very tricky for complex systems such as
this one in which allowance must be made for the possible coexistence of
closely-related phases and the presence of anisotropic peak broadening. The
choice of any particular model should take into account not only the quality
of the refinement as judged by the agreement factors and goodness-of-fit,
but also the diffraction profiles of alternative models in selected key
regions of the pattern, and, if feasible, data from complimentary structural
techniques such as electron diffraction.
\begin{center}
\textbf{ACKNOWLEDGMENTS}
\end{center}
We would like to acknowledge the
support of the NIST Center for Neutron Research, U. S. Department
of Commerce, and to thank B. H. Toby for experimental assistance
at beamline BT1. Financial support from the
U.S. Department of Energy, Division of Materials Sciences, under contract No. DE-AC02-98CH10886
is also acknowledged.

\newpage

\begin{table}[p]
\centering \caption{{\protect\small Refined structural parameters
for tetragonal PbZr$ _{0.52}$Ti$_{0.48}$O$_{3}$
 at 550 K, space group $P4mm$, lattice parameters $a = 4.0596(1), c = 4.0999(1)$ \AA . The refinement was based on a model with the Pb atoms statistically distributed among 4(d) sites at
$x, x, 0,$ corresponding to local displacements along
$\left\langle 110\right\rangle $ directions. Figures in
parentheses denote standard errors referred to the least
significant digit. $R_{wp}$,
 $R_{B}$ and $\protect\chi ^{2}$
are agreement factors as defined in Ref. \protect\onlinecite{FULLPROF}. }}
\begin{tabular}{lcccc}\\
\hline\hline
 & \small $x$ & \small $y$ & \small $z$ & {\small$U$(\AA $^{2}$)} \\
\hline
\small Pb & 0.033(1) & 0.033(1) & 0.0 & 0.028(1) \\
\small Zr/Ti & 0.5 & 0.5 & 0.450(2) & 0.005(1) \\
\small O(1) & 0.5 & 0.5 & -0.061(1) & 0.027(1) \\
\small O(2) & 0.5 & 0.0 & 0.427(1) & 0.027(1) \\
& & & & \\
\small $R_{wp}$ & \multicolumn{4}{c}{0.048} \\
\small $R_{B}$ &  \multicolumn{4}{c}{0.034} \\
\small $\chi 2$ &  \multicolumn{4}{c}{1.20} \\
 \hline\hline
\end{tabular}
\end{table}
\begin{table}[p]
\centering \caption{{\protect\small Refined structural parameters
for monoclinic PbZr$_{0.52}$Ti$_{0.48}$O$_{3}$
 at 20 K, single-phase model with space group $Ic$, for
the [001]-tilt and $y$-shift models described in the text. The Pb
atom was fixed at the origin, and the O(2) and O(3) temperature
factors were constrained to be equal. Also listed are the
parameters recently reported by Hatch $et~ al.$
\protect\cite{Hatch}, but with the values transformed from $Cc$
 to $Ic$ symmetry. $V_{0}$ and $c_{0}/a_{0}$ represent respectively
the volume and axial ratio of the primitive pseudocubic cell, with
$c_{0} =c/2$ and $a_{0}=(a+b)/2\surd 2$. }}
\begin{tabular}{lccc}\\
\hline\hline
& \multicolumn{2}{c}{\small  Present study} &  {\small Hatch \it{et al.}}\\
& {\small $[001]-$tilt} & {\small  $y-$shift} & {\small $y-$shift}
\\  \hline
{\small $a($\AA $)$} & 5.7131(1) & 5.7131(1) & 5.7312(7) \\
{\small $b($\AA $)$} & 5.7000(1) & 5.7001(1) & 5.7093(6) \\
{\small $c($\AA $)$} & 8.2679(2) & 8.2683(3) & 8.2363(7) \\
{\small $\beta (^{o})$} & 90.475(2) & 90.473(2) & 90.50(1) \\
{\small $V_{o}($\AA $^{3})$} & 67.31 & 67.31 & 67.37 \\
{\small $c_{o}/a_{o}$} & 1.0245 & 1.0246 & 1.0181 \\
& & & \\
{\small Pb: \ \  $U($\AA $^{2})$} & 0.013(1) & 0.012(1) & 0.013(1) \\
{\small Zr/Ti: \ \ \ \  $x$} & 0.524(2) & 0.524(2) & 0.519(5) \\
{\small \ \ \ \ \ \ \ \ \ \ \ \ \ \ $z$} & 0.219(1) & 0.218(1) & 0.216(2) \\
{\small \ \ \ \ \ \ \ \ $U($\AA $^{2})$} & 0.002(1) & 0.003(2) & 0.006(4) \\
{\small O(1): \ \ \ \ \ \ $x$} & 0.542(1) & 0.543(1) & 0.548(3) \\
{\small \ \ \ \ \ \ \ \ \ \ \ \ \ \ $z$} & -0.046(1) & -0.046(1) & -0.044(1) \\
{\small \ \ \ \ \ \ \ \  $U($\AA $^{2})$} & 0.011(1) & 0.011(1) & 0.011(3) \\
{\small O(2): \ \ \ \ \ \ $x$} & 0.275(1) & 0.287(1) & 0.289(2) \\
{\small \ \ \ \ \ \ \ \ \ \ \ \ \ \ $y$} & 0.243(1) & 0.233(1) & 0.233(1) \\
{\small \ \ \ \ \ \ \ \ \ \ \ \ \ \ $z$} & 0.193(1) & 0.194(1) & 0.196(1) \\
{\small \ \ \ \ \ \ \ \ $U($\AA $^{2})$} & 0.010(1) & 0.011(1) & 0.009(1) \\
{\small O(3): \ \ \ \ \ \ $x$} & 0.801(1) & 0.787(1) & 0.789(2) \\
{\small  \ \ \ \ \ \ \ \ \ \ \ \ \ \ $y$} & 0.768(1) & 0.767(1) & 0.767(1) \\
{\small  \ \ \ \ \ \ \ \ \ \ \ \ \ \ $z$} & 0.193(1) & 0.194(1) & 0.196(1) \\
{\small  \ \ \ \ \ \ \ \ $U($\AA $^{2})$} & 0.010(1) & 0.011(1) & 0.009(1) \\
& & & \\
\small $R_{wp}$ & 0.073 & 0.076 & 0.086 \\
\small $R_{B}$ & 0.041 & 0.047 & 0.040 \\
\small $\chi ^{2}$ & 1.52 & 1.62 & 1.21\\
\hline\hline
\end{tabular}
\end{table}

\begin{table}[p]
\centering \caption{{\protect\small Refined structural parameters
for PbZr$_{0.52}$Ti$_{0.48}$O$_{3}$ at 20 K, two-phase model with
space groups $Cm$ and $R3c$. The Pb and Zr/Ti atoms
 were fixed at the origin for the $Cm$ and $R3c$ refinements respectively,
and the temperature factors for the separate atoms were
constrained to be the same in both structures. Also listed are the
10 K parameters recently reported by Frantti {\it et
al.}\protect\cite{Frantti3}}}
\begin{tabular}{lcccc}\\
\hline\hline
& \multicolumn{2}{c}{\small Present study} & \multicolumn{2}{c}{\small %
Frantii {\it et al.}} \\
& {\small $Cm$} & {\small $R3c$} & {\small $Cm$} & {\small $R3c$} \\
\hline \\
{\small $a($ \AA $)$} & 5.7120(1) & 5.7415(6) & 5.7097(7) & 5.744(2) \\
{\small $b($ \AA $)$} & 5.6988(1) & - & 5.6984(7) & - \\
{\small $c($ \AA $)$} & 4.1353(1) & 14.208(3) & 4.1367(3) & 14.212(8) \\
{\small $\beta (^{o})$} & 90.479(2) & - & 90.449(8) & - \\
{\small $V_{o}($ \AA$^{3})$} & 67.32 & 67.60 & 67.29 & 67.68 \\
{\small $c_{o}/a_{o}$} & 1.0257 & 1.0 & 1.0256 & 1.0 \\
&  &  &  &  \\
{\small Pb:  \ \ \ \ \ \ \ \ \ \ \ $z$} & - & 0.283(4) & - &
0.282(5) \\
\ \ \ \ \ \ \ \ \ {{\small $U$(\AA }$^{2})$} & 0.012(1) & 0.012(1)
& 0.009(1)
& 0.004(6) \\
{\small Zr/Ti:  \ \ \ \ \ \ \ $x$} & 0.531(2) & - &
0.539(3) & - \\
{\small \ \ \ \ \ \ \ \ \ \ \ \ \ \ \ \ $z$} & 0.441(2) & - & 0.441(3) & - \\
\ \ \ \ \ \ \ \ \ {{\small $U$(\AA }$^{2})$} & 0.004(1) & 0.004(1)
& 0.001(2)
& 0.001(2) \\
{\small O(1): \ \ \ \ \ \ \ \ $x$} & 0.543(1) & - &
0.540(1) & - \\
\ \ \ \ \ \ \ \ \ \ \ \ \ \ \ \ {\small $ z$} & -0.090(1) & - & -0.092(2) & - \\
\ \ \ \ \ \ \ \ \ {{\small $U$(\AA }$^{2})$} & 0.008(1) & - & 0.011(2) & - \\
{\small O(2): \ \ \ \ \ \ \ \ $x$} & 0.288(1) & 0.137(3) & 0.283(1) & 0.148(3) \\
\ \ \ \ \ \ \ \ \ \ \ \ \ \ \ \ {\small $ y$} & 0.254(1) & 0.347(3) & 0.253(1) & 0.354(3) \\
\ \ \ \ \ \ \ \ \ \ \ \ \ \ \ \ {\small $ z$} & 0.389(10 & 0.081(4) & 0.388(1) & 0.081(6) \\
\ \ \ \ \ \ \ \ \ {{\small $U$(\AA }$^{2})$} & 0.015(1) & 0.015(1) & 0.013(1) & 0.007(5) \\
 &  &  &  &  \\
{$f${\small (wt fraction)}} & 0.89(1) & 0.11(1) & 0.87 & 0.13 \\
  &  &  &  &  \\
{\small $R_{wp}$} & \multicolumn{2}{c}{0.067} &
\multicolumn{2}{c}{0.058} \\
{\small $R_{B}$} & 0.039 & 0.067 & \multicolumn{2}{c}{0.042} \\
{\small $\chi ^{2}$} & \multicolumn{2}{c}{1.27} &
\multicolumn{2}{c}{2.28}\\
\hline\hline
\end{tabular}
\end{table}
\begin{table}[p]
\centering \caption{{\protect\small Refined structural parameters
for monoclinic PbZr$ _{0.52}$Ti$_{0.48}$O$_{3}$ at 20 K, two-phase
model with space groups $Cm$ and $Ic$. The atomic positions for
$Ic$ symmetry were based upon the [001]-tilt model described in
text and constrained to be equivalent to those for $Cm$ except for
one additional parameter $\protect\delta$ corresponding to O(2)
and O(3) displacements in the $x$ and $y$ directions due to
tilting. For comparison with the $Ic$ structure, the O(2) and O(3)
positions for the $Cm$ structure are shown separately, although in
fact they are symmetry-equivalent. The temperature factors for the
separate atoms were constrained to be the same in both structures.
}}
\begin{tabular}{lcc}\\
\hline\hline
& {\small $Cm$} &{\small  $Ic$} \\
$a(${\small \AA}$)$ & 5.7097(1) & 5.7401(7) \\
$b(${\small \AA}$)$ & 5.6988(1) & 5.7188(8) \\
$c(${\small \AA}$)$ & 4.1373(1) & 8.2098(11) \\
$\beta (^{o})$ & 90.473(2) & 90.550(10) \\
$V_{o}(${\small \AA }$^{3})$ & 67.31 & 67.37 \\
$c_{o}/a_{o}$ & 1.0257 & 1.0127 \\
&  &  \\
{\small Pb:  \ \ \ \    $U$(\AA$^{2})$} & 0.012(1)
& 0.012(1) \\
{\small Zr/Ti:  \ \ \ \ \ \ \  $x$} & 0.530(2) & 0.530(2)-
\\
\ \ \ \ \ \ \ \ \ \ \ \ \ \ \ \ {\small $z$} & 0.437(1) & 0.218(1) \\
\ \ \ \ \ \ \ \ \ {\small $U$(\AA$^{2})$} & 0.003(1) & 0.003(1) \\
{\small O(1): \ \ \ \ \ \ \ \ $x$} & 0.541(1) & 0.541(1)
\\
\ \ \ \ \ \ \ \ \ \ \ \ \ \ \ \ {\small $z$} & -0.089(1) & -0.045(1) \\
\ \ \ \ \ \ \ \ \ {\small $U$(\AA$^{2})$} & 0.011(1) & 0.011(1) \\
{\small O(2): \ \ \ \ \ \ \ \ $x$} & 0.286(1) & 0.257(1)
\\
\ \ \ \ \ \ \ \ \ \ \ \ \ \ \ \ {\small $y$} & 0.254(1) & 0.225(1) \\
\ \ \ \ \ \ \ \ \ \ \ \ \ \ \ \ {\small $z$} & 0.390(1) & 0.195(1) \\
\ \ \ \ \ \ \ \ \ {\small $U$(\AA$^{2})$} & 0.015(1) & 0.015(1) \\
{\small \ O(3):  \ \ \ \ \ \ \  $x$} & 0.786(1) & 0.814(1)
\\
\ \ \ \ \ \ \ \ \ \ \ \ \ \ \ \ {\small $y$} & 0.754(1) & 0.783(1) \\
\ \ \ \ \ \ \ \ \ \ \ \ \ \ \ \ {\small $z$} & 0.390(1) & 0.195(1) \\
\ \ \ \ \ \ \ \ \ {\small $U$(\AA$^{2})$} & 0.015(1) & 0.015(1) \\
&  &  \\
$f${\small (wt fraction)} & 0.78(2) & 0.22(2) \\
 &  &  \\
{\small $R_{wp}$} & \multicolumn{2}{c}{0.064} \\
{\small $R_{B}$} & 0.039 & 0.053 \\
$\chi ^{2}$ & \multicolumn{2}{c}{1.16}\\
\hline\hline
\end{tabular}
\end{table}

\begin{table}[p]
\begin{centering}
\caption{{\protect\small Refined structural parameters for the
$Cm$
 and $P4mm$ phases in PbZr$_{0.52}$Ti$_{0.48}$O$_{3}$ at 325
K with the three-phase model described in text. The temperature
factors for the separate atoms were constrained to be the same in
both structures. Also listed are the room-temperature parameters
recently reported in an x-ray study by Ragini
$et~al.$\protect\cite{Ragini2} and in a neutron study by Frantti
$et~al.$\protect\cite{Frantti4} The weight fraction of the cubic
$Pm3m$ phase was determined as 0.06. }}
\begin{tabular}{lcccccc}\\
 \hline\hline
 & \multicolumn{2}{c}{\small Present study} & \multicolumn{2}{c}{\small %
Ragini $et~al.$} & \multicolumn{2}{c}{\small Frantti $et~al.$} \\
 & {\small $Cm$} & {\small $P4mm$} & {\small $Cm$} & {\small $P4mm$} & {\small $Cm$} &
{\small $P4mm$} \\
\hline {\small $a($\AA$)$} & 5.7268(3) & 4.0393(1) & 5.7520(1) &
4.0429(2) &
5.7129(3) & 4.0550(4) \\
{\small $b($\AA$)$} & 5.7187(3) & - & 5.7431(2) & - & 5.7073(3) & - \\
{\small $c($\AA$)$} & 4.1230(2) & 4.1388(1) & 4.0912(4) &
4.1318(3) &
4.1436(1) & 4.1097(6) \\
{\small $\beta (^{o})$}  & 90.393(5) & - & 90.48(1) & - & 90.199(3) & - \\
{\small $V_{o}($\AA$^{3})$} & 67.51 & 67.53 & 67.57 & 67.53 &
67.55 & 67.58
\\
{\small $c_{o}/a_{o}$}  & 1.0189 & 1.0246 & 1.0067 & 1.0219 & 1.0262 & 1.0135 \\
& & & & & & \\
{\small Pb: \ \ \ \ \ \ \ \ \ $z$} & - & 0.035(2)$^{a}$ & - & - & - & - \\
 \ \ \ \ \ \ \ \ {\small $U($\AA$^{2})$} & 0.017(1) & 0.017(1) & 0.107$%
^{b} $ & 0.030$^{c}$ & 0.021$^{d}$ & 0.019(1) \\
{\small Zr/Ti: \ \ \ \ \ $x$} & 0.530(4) & - & 0.578(3) & - & 0.507(2)/0.494(4)$^{e}$ & - \\
\ \ \ \ \ \ \ \ \ \ \ \ \ \ \ {\small $z$} & 0.432(5) & 0.442(2) &
0.473(3) & 0.447(2) &
0.426(1)/0.404(4)$^{e}$ & 0.431(4) \\
 \ \ \ \ \ \ \ \ {\small $U($\AA$^{2})$} & 0.003(1) & 0.0043(1) & 0.015(1)
& 0.005(2) & 0.004(1) & 0.019(1) \\
{\small O(1): \ \ \ \ \ \ \ $x$} & 0.540(2) & - & 0.50(1)
& - & 0.522(1) & - \\
\ \ \ \ \ \ \ \ \ \ \ \ \ \ \ {\small $z$} & -0.080(3) & -0.085(2)
& -0.10(1) & -0.109(6)
& -0.090(1) & -0.080(2) \\
 \ \ \ \ \ \ \ \ {\small $U($\AA$^{2})$} & 0.016(1) & 0.016(1) & 0.00(1) &
0.029(1) & 0.013(1) & 0.019(1) \\
{\small $O(2):$ \ \ \ \ \ \ $x$}&  0.287(2) & - & 0.36(1)
& -- & 0.270(1) & - \\
\ \ \ \ \ \ \ \ \ \ \ \ \ \ \ {\small $y$} & 0.255(1) & - & 0.219(8) & - & 0.252(1) & - \\
\ \ \ \ \ \ \ \ \ \ \ \ \ \ \ {\small $z$} & 0.400(2) & 0.0395(1)
& 0.404(8) & 0.389(3) &
0.391(1) & 0.400(1) \\
 \ \ \ \ \ \ \ \ {\small $U($\AA$^{2})$} & 0.021(2) & 0.021(2) & 0.04(1) &
0.029(1) & 0.013(1) & 0.019(1) \\
  &  &  &  &  &  &  \\
{\small $f$ (wt fraction)}& 0.33(1) & 0.61(2) & 0.58 & 0.42 & 0.69 & 0.31 \\
 &  &  &  &  &  &  \\
{\small $R_{wp}$} & \multicolumn{2}{c}{0.051} &
\multicolumn{2}{c}{0.128} & \multicolumn{2}{c}{0.021} \\
{\small $R_{B}$} & 0.043 & 0.030 & 0.041 & 0.062 & \multicolumn{2}{c}{-} \\
{\small $\chi ^{2}$} & \multicolumn{2}{c}{1.47} &
\multicolumn{2}{c}{3.39} & \multicolumn{2}{c}{2.69}\\
\hline\hline
\end{tabular}
\end{centering}
\begin{flushleft}

{\footnotesize{\protect

(a) Pb atoms statistically distributed among 4(d) sites at $x, x,
0$.

(b) Equivalent isotropic $U$  $(U_{11}= 0.221, U_{22}= 0.027,
U_{33}= 0.074, U_{13}= 0.030$ \AA$^{2}).$

(c) Equivalent isotropic $U$ $(U_{11}= U_{22}= 0.031, U_{33}=
0.027$ \AA$^{2}$).

(d) Equivalent isotropic $U$  $(U_{11}= 0.027, U_{22}= 0.026,
U_{33}= 0.011, U_{13}= 0.013$ \AA$^{2}$).

(e) Zr and Ti parameters refined independently.}}
\end{flushleft}

\end{table}

\newpage

\begin{figure}[tbp]
\includegraphics {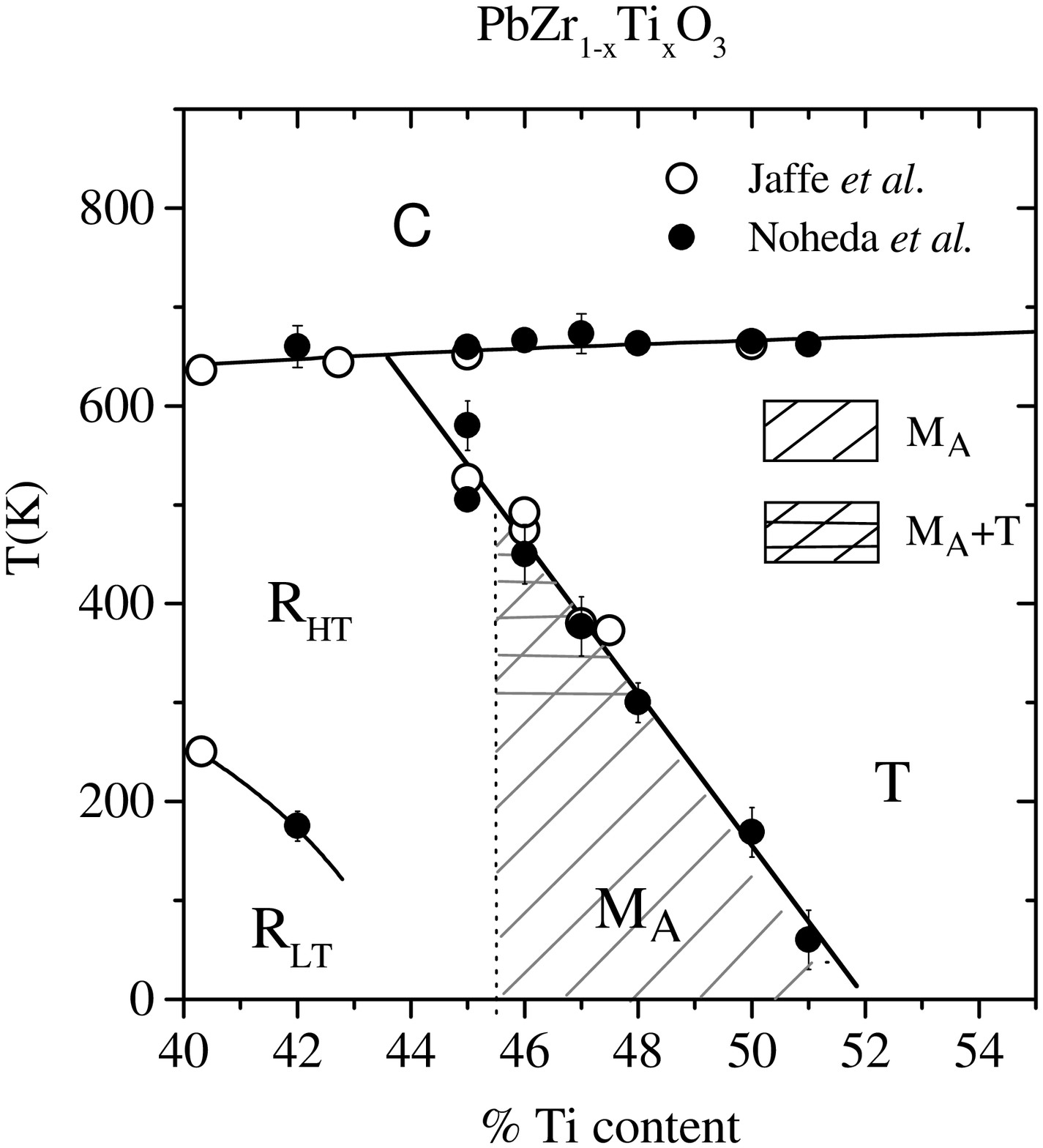}
\caption{{\protect\small PZT phase diagram as originally proposed
by Jaffe et al. in Ref. \onlinecite{Jaffe} (open circles) with the
modifications reported by Noheda et al. in Ref.
\onlinecite{Noheda3} (full circles). The
various phases described in the text are denoted by C (cubic }$Pm3m$%
{\protect\small ), R}$_{\rm HT}${\protect\small \ (rhombohedral }$R3m$%
{\protect\small), R}$_{\rm LT}${\protect\small \ (rhombohedral }$R3c$%
{\protect\small ), T (tetragonal }$P4mm${\protect\small ), and M}$_{\rm A}$%
{\protect\small \ (monoclinic} $C{m}${\protect\small ). }}
\end{figure}

\begin{figure}[tbp]
\includegraphics{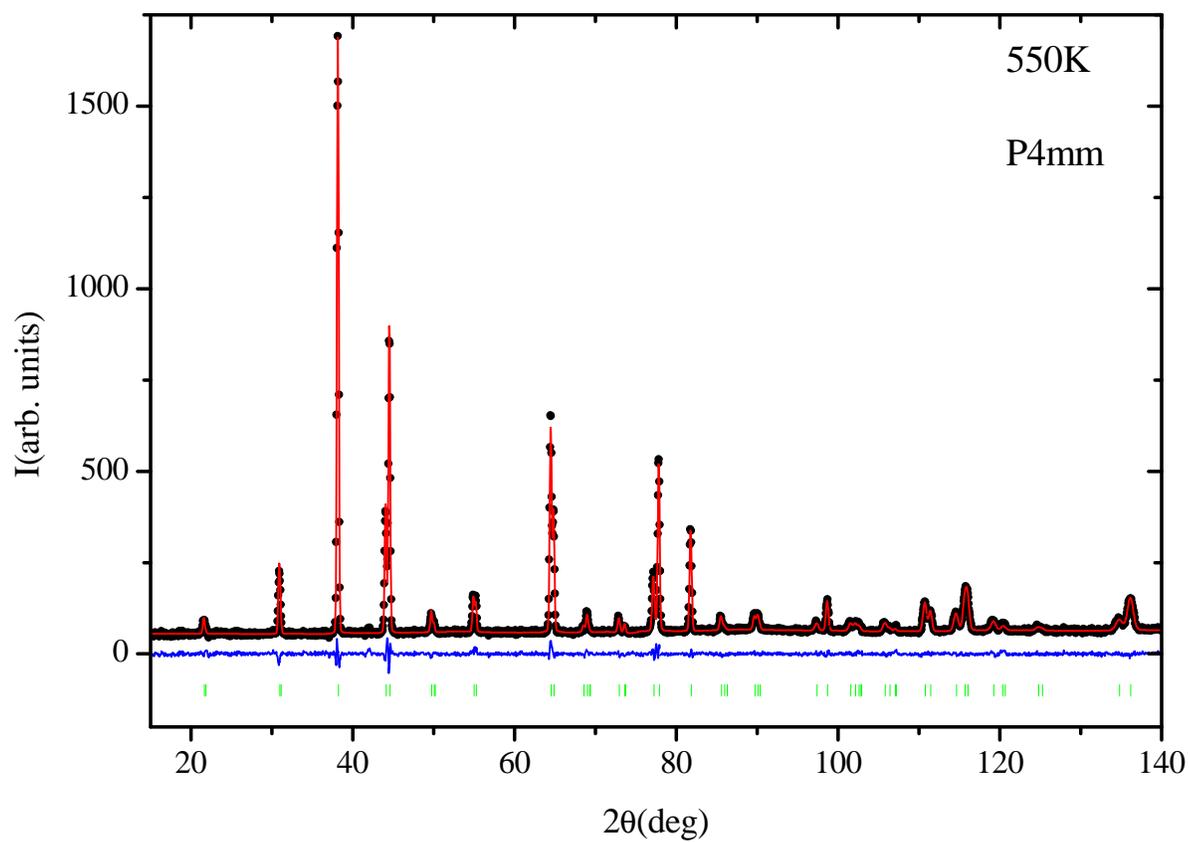}
\caption{{\protect\small (Color) Observed and calculated
diffraction profiles from
the Rietveld refinement of PbZr}$_{0.52}${\protect\small Ti}$_{0.48}$%
{\protect\small O}$_{3}${\protect\small \ at 550 K, with space group }$P4mm$%
{\protect\small . The difference plot is shown below, with  short
vertical markers denoting the calculated peak positions.}}
\end{figure}

\begin{figure}[tbp]
\includegraphics{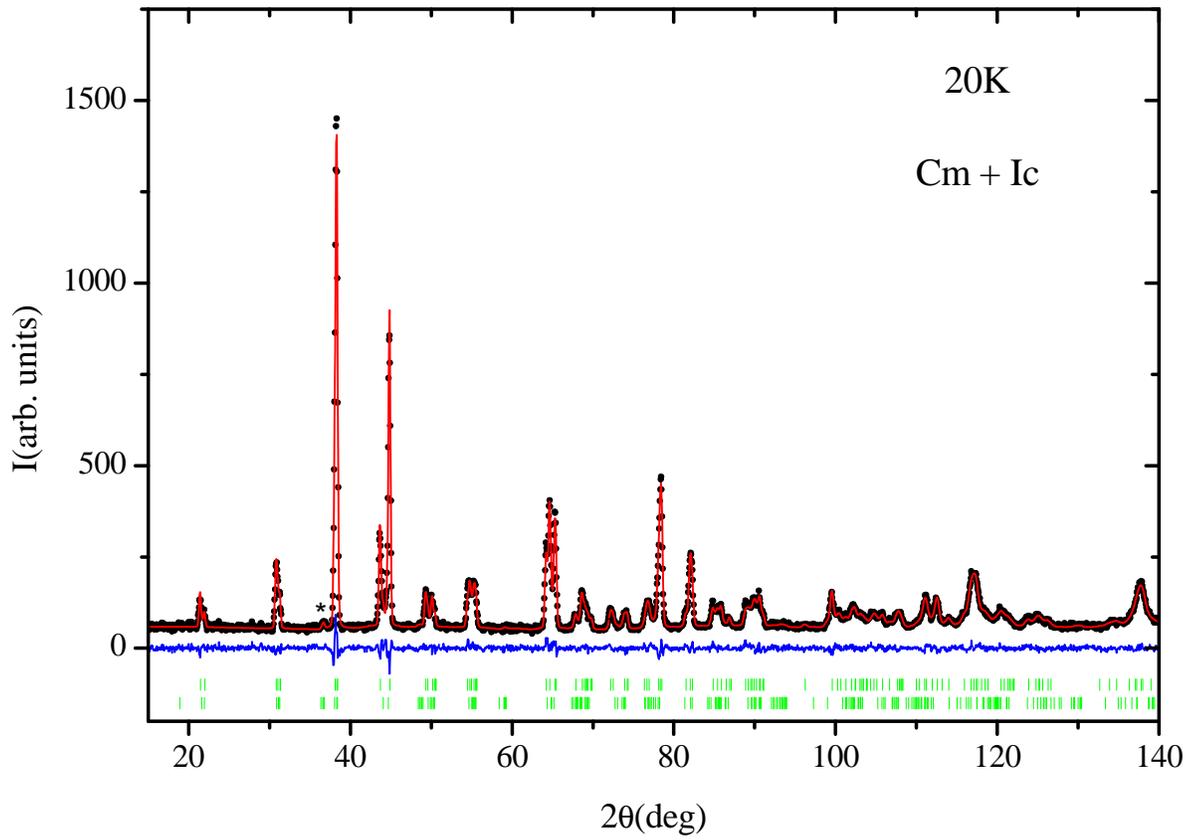}
\caption{{\protect\small (Color) Observed and calculated
diffraction profiles from
the two-phase Rietveld refinement of PbZr}$_{0.52}${\protect\small Ti}$%
_{0.48}${\protect\small O}$_{3}${\protect\small \ at 20 K, with space groups
}$Cm${\protect\small \ and }$Ic${\protect\small . The difference plot is
shown below, with  upper and lower sets of vertical markers denoting
 the calculated peak positions for }$Cm${\protect\small \ and }$Ic$%
{\protect\small \ respectively. The position of the weak superlattice peak at
2$\theta \approx 36.8^{\circ}$ (pseudocubic 3/2 1/2 1/2) is indicated with an asterisk.}}
\end{figure}

\begin{figure}[tbp]
\includegraphics{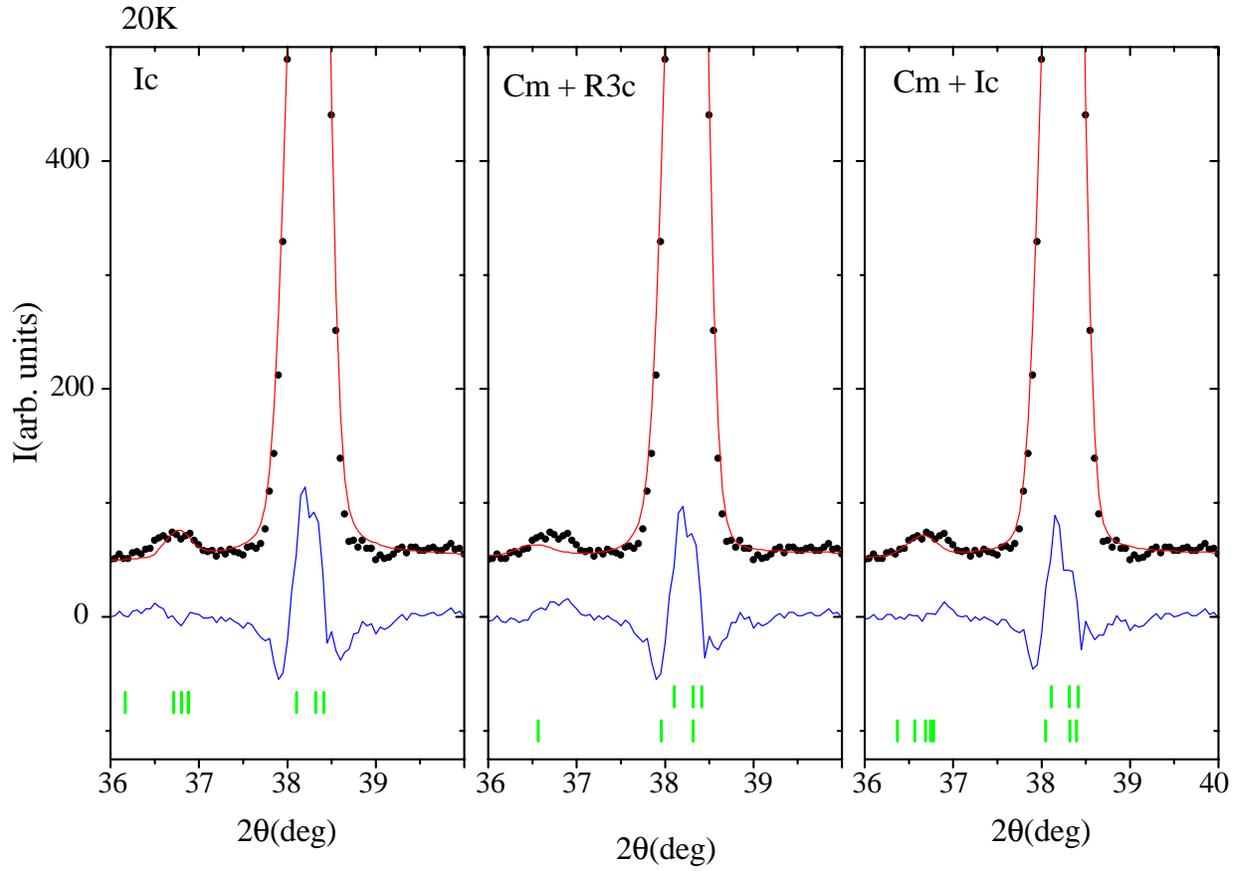}
\caption{{\protect\small (Color) Observed and calculated
diffraction profiles and difference plots in the region around the
strongest superlattice peak from
PbZr}$_{0.52}${\protect\small Ti}$_{0.48}${\protect\small O}$_{3}$%
{\protect\small \ at 20 K for single-phase }$Ic${\protect\small \ (left
panel), two-phase }$Cm+R3c${\protect\small \ (center panel), and two-phase }$%
Cm+Ic${\protect\small \ (right panel).}}
\end{figure}

\begin{figure}[tbp]
\includegraphics{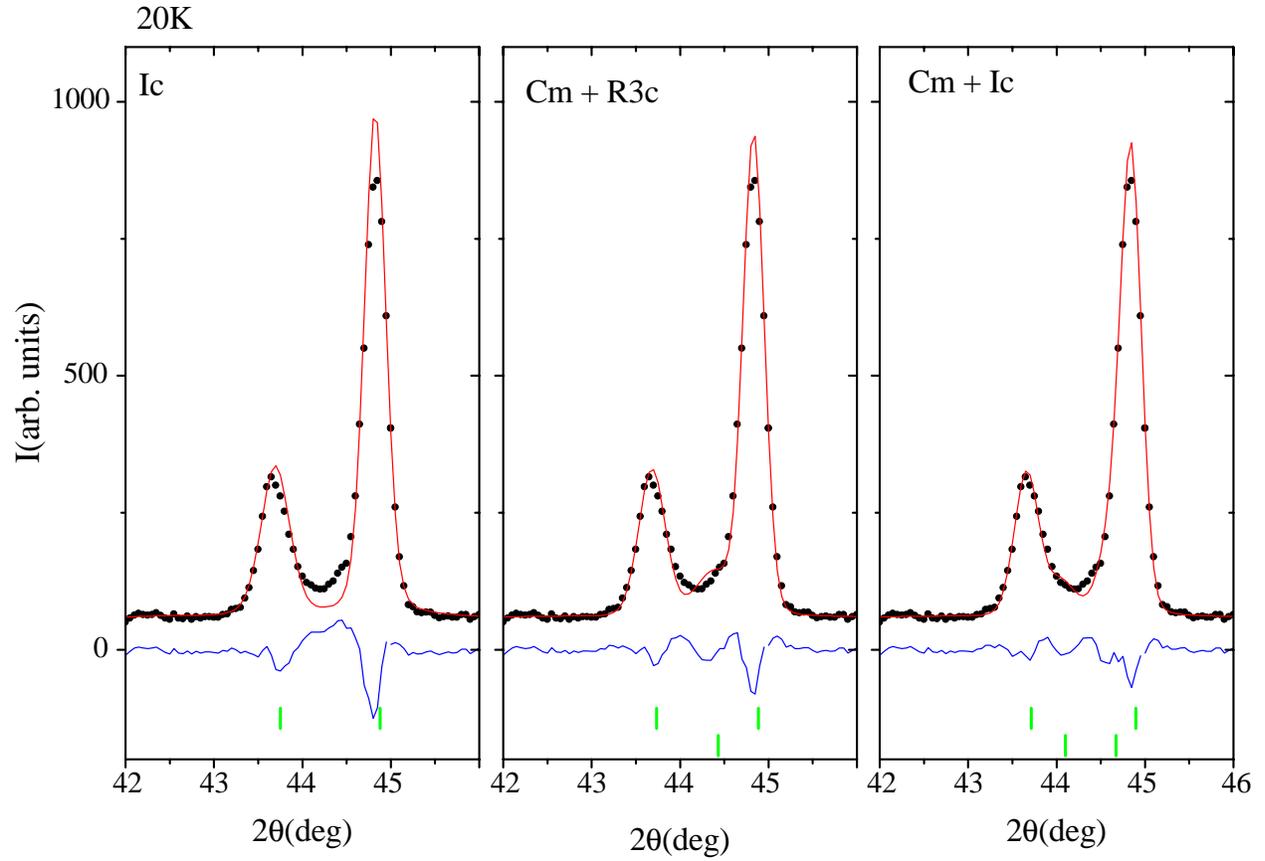}
\caption{{\protect\small (Color) Observed and calculated
diffraction profiles and difference plots in the region around the
pseudocubic (200) reflection from
PbZr}$_{0.52}${\protect\small Ti}$_{0.48}${\protect\small O}$_{3}$%
{\protect\small \ at 20 K for single-phase }$Ic${\protect\small \ (left
panel), two-phase }$Cm+R3c${\protect\small \ (center panel), and two-phase }$%
Cm+Ic${\protect\small \ (right panel).}}
\end{figure}

\begin{figure}[tbp]
\includegraphics{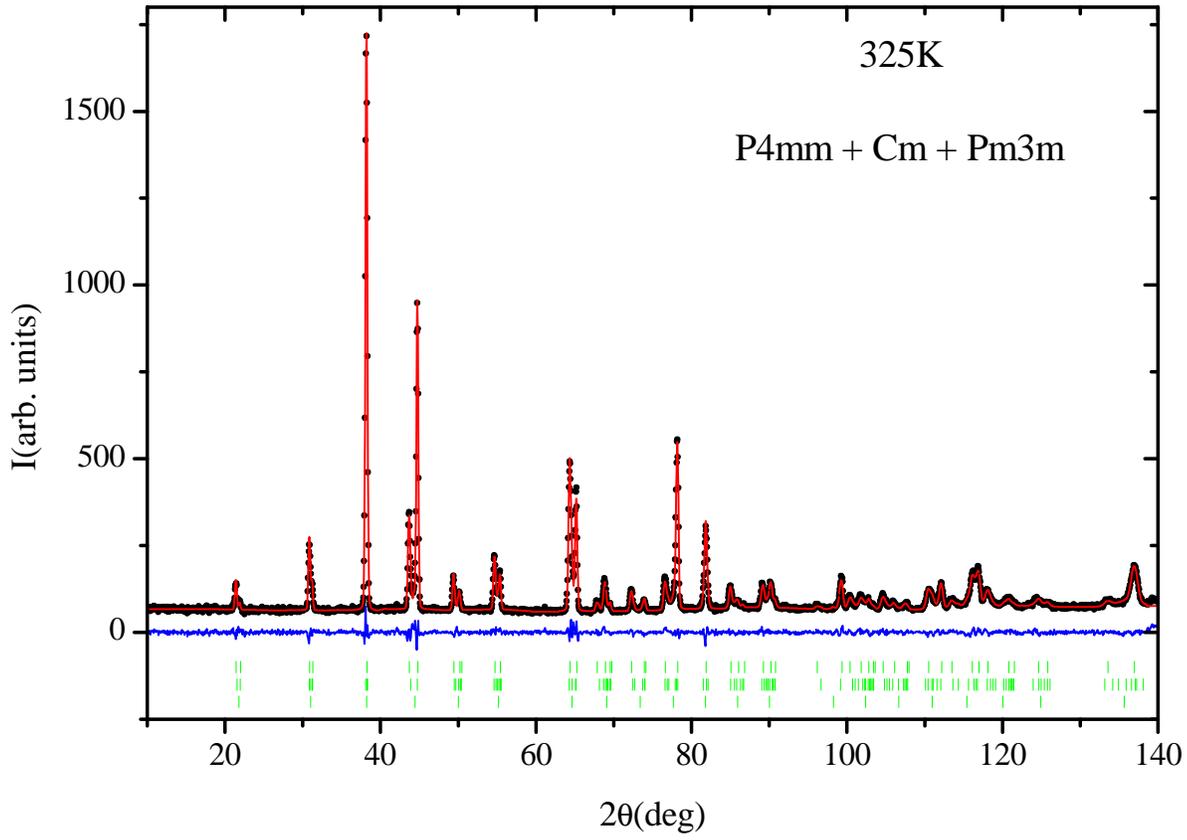}
\caption{{\protect\small (Color) Observed and calculated
diffraction profiles from
the three-phase Rietveld refinement of PbZr}$_{0.52}${\protect\small Ti}$%
_{0.48}${\protect\small O}$_{3}${\protect\small \ at 325 K, with space
groups }$P4mm${\protect\small , }$Cm${\protect\small \ and }$Pm3m$%
{\protect\small . The difference plot is shown below, with upper, middle
and lower sets of vertical markers denoting the calculated peak
positions for }$P4mm${\protect\small , }$Cm${\protect\small \ and }$Pm3m$%
{\protect\small \ respectively.}}
\end{figure}

\end{document}